\documentclass{emulateapj}	

\newcommand{\czhel}{\mbox{$cz_{\rm hel}$}}	
\newcommand{\dvlos}{\mbox{$\Delta v$}}		
\newcommand{\etal}{et al.}			
\newcommand{\hi}{\ion{H}{1}}			
\newcommand{\kms}{km~s$^{-1}$}			
\newcommand{\msun}{\mbox{${\cal M}_{\odot}$}}	
\newcommand{\n}{NGC~}				
\newcommand{\zo}{Z{$_\odot$}}			

\shorttitle{NGC~3256 Tidal Tails Clusters}
\shortauthors{Trancho et al.}

\begin{document}

\title{Gemini Spectroscopic Survey of Young  Star Clusters in Merging/Interacting Galaxies. I. NGC~3256 Tidal-Tail Clusters}  


\author{Gelys Trancho}
\affil{Universidad de La Laguna, Tenerife, Canary Island, Spain}
\affil{Gemini Observatory, 670 N. A'ohoku Place, Hilo, HI 96720, USA}
\email{gtrancho@gemini.edu}
\author{Nate Bastian}
\affil{Department of Physics and Astronomy, University College London, Gower Street, London, WC1E 6BT, United Kingdom}
\author{Fran\c{c}ois  Schweizer}
\affil{Carnegie Observatories, 813 Santa Barbara Street,  Pasadena CA 91101-1292, USA}
\and
\author{Bryan W. Miller}
\affil{Gemini Observatory, Casilla 603, La Serena, Chile}


\begin{abstract}

We present  Gemini optical spectroscopy of three young star
  clusters  in the western tidal tail of  \n3256. Compact star clusters (as opposed to dwarf-galaxy candidates) in tidal tails are rare, with these three clusters the first for which detailed quantitative spectroscopy has ever been obtained. We find that two of these clusters
   appear to be coeval, while the third is approximately two times
  older ($\sim$200 Myr vs. $\sim$80 Myr).  All three clusters are
   massive (1--$3\times 10^5 \msun $) and appear to be of roughly solar
  metallicity.  Additionally, the three
  clusters appear to be relatively large (R$_{\rm eff} = 10-20$~pc),
  possibly reflecting weak compression at the time of formation and/or the weak influence of the tidal field of the galaxy. All three clusters have velocities consistent with the general 
  trend of the \ion{H}{1} velocities in the tidal tail. We conclude that if the loosely 
   bound tail material of NGC~3256 gets stripped during future interactions of this galaxy within its  group,  these three clusters may become part of the intra-group medium.
  
\end{abstract}


\section{INTRODUCTION}

Massive star clusters are usually considered to be components of
galaxies, however there is a growing body of evidence that there are
also populations of star clusters that may be bound to the gravitational
potential of galaxy groups and clusters rather than to individual
galaxies \citep{west95, kp99, bassino03, jordan03}.  As with the
intracluster stars \citep{ferguson98, mijos05, krick05} and PNe
\citep{aguerri05} it is thought that these star clusters  are mostly old
objects stripped from galaxies as they interact with other galaxies in their group or  cluster \citep{murante04, yahagi05}.  Alternatively, {\it
young} star clusters formed during galaxy mergers and
interactions may also get flung into the intracluster environment.
This paper investigates the plausiblility of this process using
spectroscopy of young star clusters in the tidal tails of \n3256.

Massive star clusters are formed wherever high gas pressures cause
localized star formation efficiencies to exceed 20\%--50\% \citep{jog,ee97,
mclaughlin99}.\\
Bound star clusters may form whenever the local SFE is above
  30\% (e.g. Goodwin 1997; Boily \& Kroupa 2003 and references
  therein).
  Imaging with the Hubble Space Telescope has revealed
that young star clusters are formed copiously in the central parts of
galaxy mergers, strengthening theories in which giant elliptical
galaxies are formed by the merger of spirals (Ashman \& Zepf 1992;
Whitmore et al.~1993; Miller et al.~1997; Zepf et al.~1999).  Young
star cluster candidates have also been found in the tidal tails of
merging and interacting galaxies \citep{knier03, tran03, degrijs03,
bastian05}.  Photometric ages suggest that these clusters were formed
in the tails, but spectroscopic ages and kinematics are needed to
confirm this. Simulations show that the majority of the tails of
mergers will probably fall back onto the merger remnant \citep{hm95}.
However, the outer parts of the tails become unbound and that material will
become part of the intergalactic medium.  Therefore, a mechanism
exists to inject younger stellar populations in to intracluster
regions but kinematics of the young clusters in tidal tails are needed
to determine in what instances this occurs.

\n3256 is probably one of the best nearby systems for studying the properties
of young star clusters in tidal tails.  It is a classic merging
system, exhibiting a double nucleus \citep{engl03a} , a confusion of dust lanes
in the central regions, loops and shells surrounding the main body,
and two tidal tails that extend up to $\sim50$~kpc in \hi  \citep{engl03b}. 
 \citet{toomre77} classified it as intermediate in merger stage
between \n4038/39 and \n7252.  However, it is exceptional in having
the highest molecular gas mass, far-infrared luminosity,
and X-ray luminosity, and  star-formation
rate of the mergers in the Toomre sequence \citep{zepf99}.  
Unlike in the Antennae, where they found no cluster candidates in the tidal tails, \citet{knier03} found a
significant excess of star cluster candidates in the tidal tails of
\n3256. The western tail, in particular, has a density of cluster
candidates much higher than in the tails of other merger remnants that
have been studied.  These cluster candidates have $V$ magnitudes
between 22 and 26, therefore the brightest can be studied
spectroscopically with 8-m class telescopes.

In this paper we present the first results of a large spectroscopic
survey of star clusters in galactic mergers/interactions.  We focus on
three bright star cluster candidates in the western tidal tail of \n3256
with GMOS on Gemini South.  Other targets in our survey, which will be
presented in later works, include the main body of \n3256,
NGC~4038/39 (the Antennae), \n6872, Stephan's Quintet, and M82.
The current paper is organized as follows: the  
observations are described in \S~\ref{sec:obs}.  We derive the age, mass
metallicity, and line-of-sight velocity for each of the clusters in
\S~\ref{sec:results}.  Finally,
we discuss and summarize the results in \S~\ref{sec:disc}.

\n3256 is located at $\alpha_{\rm J2000}=10^{\rm h}27^{\rm m}51\fs3$,
$\delta_{{\rm J}2000}=-43\degr54\arcmin14\arcsec$ and has a recession
velocity relative to the Local Group of $cz_{_{\rm LG}} = +2527\pm 3$
\kms, which places it at a distance of 36.1 Mpc for $H_0 = 70$ \kms\
Mpc$^{-1}$.  At that distance, adopted throughout the present paper,
$1\arcsec = 175$ pc.  The corresponding true distance modulus is
$(m-M)_0 = 32.79$. The Milky Way foreground extinction is relatively
high, $A_V=0.403$ (Schlegel et al.~1998), whence the apparent visual
distance modulus is $V-M_V = 33.19$.

\section{OBSERVATIONS AND REDUCTIONS}\label{sec:obs}

Imaging and spectroscopy of star cluster in \n3256 were obtained with
GMOS-S in semesters 2003A and 2004A. The data were obtained as part of two Director's Discretionary Time program GS-2003A-DD-1  and GS-2004A-DD-3.  The imaging covers the typical GMOS-S field, which is
approximately  $5\farcm5\times  5\farcm5$. Imaging was obtained through two
filters g' and r'. Four GMOS masks were used for the spectroscopy. We
used the B600 grating and a slit of $0\farcs75$, resulting in a
instrumental resolution of 110 km/s at 5100 \AA. The spectroscopic
observations were obtained as 8 individual exposures with exposure
time of 3600 sec each. Spectroscopy of 70 candidates yielded only 50 that
were star clusters and only three were located in the tidal tails. In
this paper we will focus on these three star clusters.  Table~1  lists these 3 star clusters. Column (1) gives the
adopted cluster ID, columns (2)-(3) the coordinates, columns (6) and
(7) g' and r' magnitudes and errors. The magnitudes have been corrected for Galactic extinction, but not for any internal extinction.

Figure 1 shows the g'  image of \n3256  (body and western tail) with the observed candidate clusters T1127, T1165 and T1149.

\begin{figure}
   \epsscale{1.0}
    \caption{Position (green square box) of the three star clusters in the Western Tail in \n3256.}
\end{figure}

The basic reductions of the data were done using a combination of the
Gemini IRAF package and custom reduction techniques. The Gemini IRAF
package is an external package built on to the core IRAF. 
All science images were bias-subtracted and then
flat-fielded with Gcal flats which were co-added
and normalized. Spectra were  extracted and wavelength calibrated with solutions
obtained from the arc exposures. The spectra
were then combined with cosmic-ray rejection, and flux
calibrated using the response function derived from our
flux standard.
In all cases, we also computed an error spectrum as the
square root of the variance of the sum spectrum.  To achieve
this, the iraf task `apall' weights each pixel in the sum by
an estimated variance based on a spectrum model and detector
noise parameters.  The error spectrum is then used to facilitate
accurate estimation of uncertainties in our analysis.

More details of the
 reductions will be described in Trancho et al (2007a), hereafter
Paper II.  Figure~2 displays the flux
 calibrated spectra of T1227, T1149, T1165 versus the observed wavelength.
  
\section{RESULTS}\label{sec:results}

\subsection{WFPC2 Sizes}
\label{sec:sizes}

In order to determine the structural parameters of the clusters in
this work, we have used archival {\it F555W-WFPC2/HST} images.  The
observations are presented in detail in Knierman et al.~(2003).  
Sizes were measured with the {\it ISHAPE} routine of Larsen (1999),
 which yielded half-light radii $R_{\rm eff}$. We fitted different
 profiles (King30, King100, Moffat15 and Moffat25) of multiple
sizes. The PSF was generated using {\it TinyTim} (Krist \& Hook~1997)
at the exact position of the clusters on the WFPC2 chips. The mean $R_{\rm eff}$
of the above profiles are given in Table~1.

Each of the three clusters is
larger than the nominal resolution limit for ISHAPE (FWHM~$\sim0.2$~pixels;
Larsen~2004), which at the assumed distance of NGC~3256 corresponds to
a linear distance of 3.2~pc.  These clusters are quite extended
(average $R_{\rm eff}$ of a Milky-Way GC or YMC  is 3$ - $4~pc, whereas the three clusters
presented here have effective radii between $\sim10-20$~pc) but still
within the range for
young clusters (e.g.~Maraston et al.~2004; Larsen~2004). 
These relatively large sizes may reflect weak compression of the GMC at
the time of cluster formation (Jog \& Solomon 1992, Elmegreen \& Efremov
1997). Additionally, since the clusters lie outside the main body of the galaxy 
their outer envelopes may be not as efficiently stripped by the galactic potential, leaving them rather large (cf. Schweizer 2004). 

\subsection{ Cluster Properties}

Using the age and metallicity sensitive spectral features, we can
determine the properties of the clusters assuming single-burst stellar
populations (e.g. Schweizer \& Seitzer 1998). However, there are
significant differences in resolution, $\alpha$/Fe-enhancement and range of ages between the
model spectra from various groups (e.g.~Bruzual \& Charlot~2003
[hereafter BC03], Vazdekis~1999 [hereafter Vaz99], Thomas \etal~2004
[hereafter TMB04]).  Whereas some models like BC03 have the
necessary resolution (3 \AA) to match the observations (4.4 \AA),
others like Vazdekis have twice the resolution (1.8 \AA), but they do not extend
to very young ages. Finally, others like TMB04 have only half the
resolution of our observations but these models take into account the effects of changing element 
abundance ratios on Lick indices, hence give Lick indices not only as a function of age and metallicity, but also as a function of the [$\alpha$/Fe] ratio. 
 
Therefore, we adopt the following method: 
First, we use the TMB04 models to confirm that at younger ages
the  $\alpha$/Fe-enhancement does not significantly affect age
determinations and therefore it can be ignored in the present
study. Then we calculate the extinction of our spectra and we use the BC03 models, due to their higher spectral
resolution, to measure ages and metallicities calculated from the output spectra using the Penalized Pixel-Fitting 
method (pPXF;  \cite{2004PASP..116..138C}). 
The pPXF method is based on the Bounded-Variables Least-Squares algorithm.
It considers the Line-Of-Sight Velocity Dispersion (LOSVD) of the stars in a cluster as a 
Gauss-Hermite series and attempts to recover it using a maximum 
penalized likelihood formalism while working in pixel 
space. It has the advantage of being robust even when the data have low signal-to-noise ratio (S/N) or when the observed LOSVD is not well 
sampled. In conclusion, the pPXF method, as realized in an IDL routine, takes in model spectra of different ages/metallicities and
weights them in order to create a best fit stellar spectrum.

 \begin{figure}
    \epsscale{0.80}
     \plotone{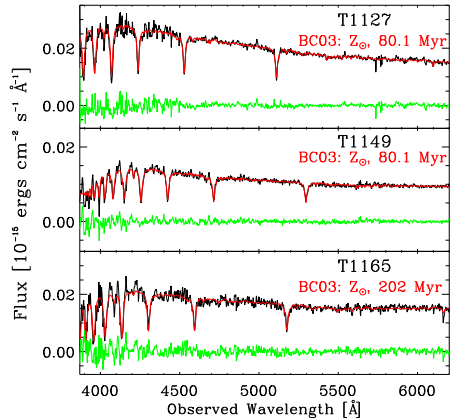}
      \caption{Optical spectra of the three young massive clusters  T1127, T1165, T1149 (black lines). Overplotted  are the model spectra (red lines) for clusters of  metallicity \zo at different ages by Bruzual \& Charlot (2003). We also show  the residuals (green lines) in the sense "observed-model".}
 \end{figure}

\subsubsection{[$\alpha$/Fe] enrichment check}
We used the TMB04 models to check and confirm that at younger ages (less that 500 Myr)  the
[$\alpha$/Fe] enrichment does not affect the age determination significantly. Especially when
  considering our typical error bars (See Figure 3) we cannot
  differentiate between the various [$\alpha$/Fe] enrichment models. 
Figure 3 shows that the [MgFe] vs age diagnostic diagram is  relatively
free from the age/metallicity degeneracy at younger
ages. Iso-[$\alpha$/Fe] tracks for three different ratios (0.0, 0.3,
and 0.5 dex) are plotted. 
  
\begin{figure}
     \epsscale{0.80}
     \plotone{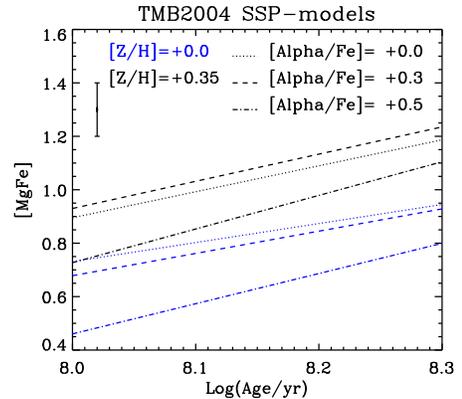}
      \caption{Lick indices [MgFe] versus age in model cluster spectra
      with age TMB04 at different [$\alpha$/Fe] enrichments. A
      typical 1$\sigma$ error bar is shown for reference.  Clearly, we
      cannot differentiate between the various models.}
 \end{figure}
   
\subsubsection{Cluster Metallicities, ages, and masses}

In order to estimate the ages of the three tidal tail star clusters,
we measured Lick line-strength indices (Faber et al.~1985;
Gonz\'alez~1993; Trager et al.~1998) and the indices defined by \cite{schweizer98}  (i.e. HHe, K, H8)  from the output spectra using the Penalized Pixel-Fitting 
method (pPXF) (In  Paper~II, we will discuss the full details of the fitting method and treatment 
of errors).
The measurements of the indices were carried out with the task INDEX (Cardiel et
al.~1998).   The errors on the measurements include the influence
  of photon-statistics and uncertainties in the radial velocities  and continuum level.  

\begin{figure}
     \epsscale{0.80}
     \plotone{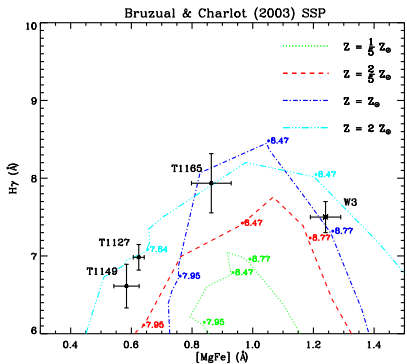}
      \caption{Determination of ages and metallicities of the
     clusters.  H$\gamma$ vs. [MgFe] from the BC03 SSP models for
     four different metallicities are shown with the measures and
     errors ($1\sigma$) form the cluster spectra are shown. }
   \label{fig:mgfe}
\end{figure}

As a first step in determining the ages and metallicities we plot the
[MgFe] ratio vs. H$\gamma$ of the BC03 models for four different
metallicities and all available ages in Fig.~\ref{fig:mgfe}. The
metallicity index [MgFe] is defined through [MgFe] =[Mgb $\times$ (0.72 $\times$
  Fe5270 + 0.28 $\times$ Fe5335)]$^{1/2}$ \citep{2003MNRAS.343..279T} , where 
Mgb, Fe5270, and Fe5335 are Lick indices expressed in Angstroms. We then
over plot the observed clusters and their $1\sigma$ error bars.

This diagram yields cluster metallicities of $2/5Z_{\odot}$ -- $2.5 Z_{\odot}$ ($\sim$2 sigma range) for T1127, T1149, and T1165. 
We also plot W3, a massive young 
cluster in the merger remnant NGC 7252, for comparison.  W3 has an age of 
~500 Myr and a metallicity of $1.0~Z_{\odot}$ (Schweizer \& Seitzer 1998). 

\begin{figure}
     \epsscale{1.0}
     \plotone{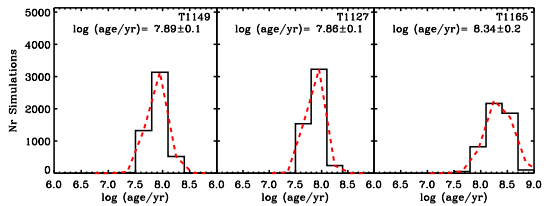}
     \caption{Histogram of the ages derived by simulating the
     effect of the errors on the age fitting routine.  See the text
     for the details of the simulations.  Note that T1127 and T1149
     appear to be coeval  while T1165 appears to be
     significantly ($\sim2.3$~times) older than the other two.}
      \label{fig:ages}
 \end{figure}

However the absorption-line indices shown in Fig.~\ref{fig:mgfe} are just one of various possible combinations 
of  line indices which we have measurements for. In
order to fully exploit the observations we compare all seven indices shown in Table~2
to the BC03 models weighted by their respective errors in a least
$\chi^2$ sense. The values of the indices are given in
Table~2.   In order to test how robust the ages and
metallicities derived are we perform the following test.  We add
random errors to each measured line index, where the added
errors are selected from a normal distribution with a standard
deviation of the observed error.  We then find the best fitting age
and metallicity, and repeat this process 5000 times for each cluster.
Finally, we fit a Gaussian to the (logarithmic) age distribution to
get the peak and standard deviation.  The results for the age
distribution are shown in Fig.~\ref{fig:ages}.  If the $3\sigma$
errors are used the results stay the same but the standard deviation
increases by about a factor 3.

As an additional check on the derived ages we also fit the full
  spectral shape of the observed clusters to the BC03 models including
  extinction.  We apply the standard galactic extinction law (with
  $A_{V}$ ranging from 0 to 10 mag) to the BC03 models in steps of
  $\Delta A_{V}=0.1$ and compare each model (24 age and 100
  extinction steps) to the observed cluster spectra in a $\chi^2$
  sense.  The fits were carried out between 4000 and 5500~\AA.  In all
  three cases presented here, the ages determined for the clusters
  through spectral indices were in excellent agreement with the results
  of fitting the spectral shape, giving us further confidence in the
  accuracy of the derived ages.  The derived extinctions are shown in Table~1.

The metallicities are also derived from the indices fitting test.  We take
the above simulations and take the mean metallicity weighted by the
number of simulations which gave each metallicity as the best fit.
Doing this we find metallicities of $1.3\pm0.7$, $1.8\pm0.6$, and
$1.4\pm0.6$ solar for T1127, T1149, and T1165 respectively.

   Once the ages are known we can derive the photometric masses of
  the star clusters by comparing their g'-band magnitude to that of
  the BC03 SSP models of the appropriate age.  The derived masses are
  given in Table~2, where we have assumed a Chabrier~(2003)
  initial mass function (IMF) for the clusters. The derived masses would be
  approximately 1.5 times higher if one assumes a Salpeter~(1955)-type
  IMF.  All three clusters appear to be quite massive, being 
  very close to the average mass of globular clusters in the Galaxy.
  As mentioned in \S~\ref{sec:sizes}, the fact that the clusters lie
  outside the main body of the galaxy suggests that they will not be
  significantly affected by the tidal field of the galaxy.  Thus,
  these clusters are not expected to loose a large amount of their
  mass due to tidal effects, so their mass loss will be dominated by
  stellar evolution and two-body relaxation.

\subsection{Cluster Velocities}
\label{sec:vel}

Table~3 gives the heliocentric radial velocities \czhel\
measured for the three star clusters.  These velocities represent
averages of velocities measured by two different methods.  First, we
determined a mean velocity from 6\,--\,7 absorption lines (H$\beta$
through H8, Ca K, and Na D) measured individually in each cluster
spectrum. And second, we also determined a mean velocity for each
cluster via cross-correlation.  As a template we used the spectra of
three radial-velocity standards observed with the same
resolution (HD~100953, HD~126248 and  HD~133955). Results from the two methods agreed to within the combined
errors, and the velocities given in Table~3 represent the
straight average of the two mean velocities measured for each cluster.

Table~3 also gives the projected distance $r_{\rm p}$ of
each cluster from the galaxy center and the cluster velocity \dvlos\
relative to the nucleus of \n3256.  This relative line-of-sight (LOS)
velocity was computed from
$$\dvlos = (cz_{\rm hel} - cz_{\rm hel,sys})/(1+z_{\rm hel,sys}),$$
where the denominator is a relativistic correction and the systemic
velocity of \n3256 is $cz_{\rm hel,sys} = 2818 \pm 3 $ \kms.  The latter
velocity is the straight average of four high-quality values published in
the literature: $2820 \pm 15$ (Feast \& Robertson 1978), $2817 \pm 10$
(L\'{\i}pari et al.~2000), and $2820 \pm 11$ \kms\ (English \& Freeman~2003)
from H$\alpha$-velocity mapping, and $2814 \pm 11$ \kms\ from the
\hi\ velocity profile (Koribalski et al.~2004).

The values of \dvlos\ for the three clusters lie in the range $-154
\la \dvlos \la -67$ \kms, indicating that all three clusters are
physically associated with \n3256 and, specifically, with its western
(W) tail.  Neutral hydrogen observations with the Australia Telescope
Compact Array show that this tail as a whole appears blue-shifted
relative to the center of the galaxy (English et al.~2003, esp.~Fig.~8).  The optical \dvlos\ of Cluster T1127 agrees with
the local \hi\ velocity in the tail to within the combined errors. However
clusters T1149 and T1165 both feature optical LOS velocities $\sim$80
\kms\ lower than the local \hi\ velocity.  This may simply reflect the
coarse resolution and heavy smoothing of the \hi\ velocity map as the
mean diameter of the synthesized beam is 23\arcsec.  {\it If\,}
these young clusters are associated with the \hi\ gas, then their
relative LOS velocities \dvlos\ indicate that the approach velocity of
the W tail is nearly twice as large (about $-$150 \kms) as previously
inferred from the \hi.  An alternative possibility, however, is that
the clusters have already decoupled kinematically from the local
gas.

The main reason why clusters eventually decouple from the surrounding
gas is that the gas experiences, and reacts to, pressure forces in
addition to gravity, while clusters orbit subject only to gravity.
(Of course, gravity fields fluctuate during a merger, but that alone
would not separate gas from stars, since both feel the same gravity.)

The most distant of the three clusters, T1149, moves with a relative
LOS velocity of $\dvlos = -154 \pm 16$ \kms.  Assuming for a moment
that its transverse motion is of similar magnitude, this $\sim$80 Myr
old cluster may have moved $\sim$11 kpc in the plane of the sky since
its birth, or nearly half of its present-day projected distance from
the center.  This allows the possibility that this cluster, and
perhaps also the other two tail clusters, may have formed
significantly closer to the center of the galactic system, presumably
as part of the tail-ejection process during the first close encounter
of the two participating disks.  If so, the cluster is likely to
follow a highly eccentric orbit within the merger remnant.

However, we emphasize that in the absence of any detailed spatial and
kinematic model for \n3256, the unknown geometry of the W tail prevents us
from drawing any firm conclusion about the likely true spatial motions of
the three clusters.  Depending on whether the clusters lie behind or in
front of the sky plane through the center of \n3256, their velocities
\dvlos\ may contain a component directed toward or away from the galaxy's
center.

\section{Summary and Discussion}\label{sec:disc}

We have studied the ages, metallicities, and [$\alpha$/Fe] ratios
of three tidal tail cluster in \n3256 based on the Lick index system.
The main results are:

\begin{enumerate}
\item T1149 and T1127 appear to be coeval, although T1165  is ~2.3 times older 
than those two ($\sim$200 Myr vs. 80 Myr).  T1165, T1127 and T1149  ages are 
consistent with being formed in the 
tail. Therefore it appears that cluster formation can proceed over
extended periods of time in tidal tails (i.e. tidal structures may not
be coeval).  This is consistent with the findings of Bastian et
al.~(2005) for the star clusters in the tidal tails of NGC~6872.

\item  All  clusters (T1127, T1149 and T1165)  are consistent with solar metallicity to within 1$\sigma$.

\item All three clusters appear to be quite massive, with masses
  greater than $10^5~M_{\odot}$, similar to the average mass of
  Galactic globular clusters.  Due to the position of the clusters
  outside the main body of NGC~3256, the clusters are not expected to
  loose a significant amount of their mass due to interactions with
  the tidal field of the galaxy.  

\item We find also that these clusters are quite extended,with
  half-light radii of $\sim10-20$~pc but still within the range for young
  clusters (e.g.~Maraston et al.~2004; Larsen~2004).  These large
  sizes may reflect weak compression at the time of formation and/or the weak influence of the tidal field of the galaxy.

\item All three clusters have velocities consistent with the general
  trend of the velocity structure of the tail. However, only T1127
  (the closest to the galaxy center) has a velocity consistent with
  the local HI velocity, while T1149 and T1165 have velocities
  $\sim80$~km/s lower than the HI.  This could be due to T1149 and
  T1165 being kinematically decoupled from the surrounding gas or
  possibly to the lower spatial resolution of the HI data.

All this is in agreement with the  dynamical time since the beginning of the merger, which is $\sim500$ Myr (English et
al. 2003).
Finally we conclude that  if the loosely bound tail material gets stripped during future interactions in the group, 
 these three clusters may well become part of the intra-group medium. 

\end{enumerate}

\begin{acknowledgements}
G.T. would like to  thank to Matt Mountain and Phil Puxley for the tremendous support throughout this project.

Based on observations obtained at the Gemini Observatory, which is operated by the
Association of Universities for Research in Astronomy, Inc., under a cooperative agreement
with the NSF on behalf of the Gemini partnership: the National Science Foundation (United
States), the Particle Physics and Astronomy Research Council (United Kingdom), the
National Research Council (Canada), CONICYT (Chile), the Australian Research Council
(Australia), CNPq (Brazil) and CONICET (Argentina)
\end{acknowledgements}


\begin{deluxetable}{lccccccc}
\def\psn{\phs\phn}
\def\pnn{\phn\phn}
\tablecolumns{8}
\tablewidth{0pt}
\tablecaption{Properties of the tidal-tail clusters. (The magnitudes have been only corrected for Galactic extinction but not for internal extinction).}
\tablehead{
\colhead{ID }&\colhead{ RA } &\colhead{ Dec}   &\colhead{ g'}  & \colhead{ r' } & \colhead{ A$_V$ }   & \colhead{ Mass} & \colhead{R$_{\rm eff}$\tablenotemark{a} } \\
 \colhead{}   &  \colhead{  (J2000)     }   &\colhead{    (J2000)}      &\colhead{   (mag)    }  &   \colhead{(mag)}  &   \colhead{(mag)}      &  \colhead{($10^{5}$  M$_{\odot}$)} & \colhead{(pc)}
}
\startdata
T1127 & 10:27:38.50    & -43:54:19.9    &     22.2$\pm$0.1 & 22.0$\pm$0.1   &    0.3 &    1.5$\pm$0.2  & 10.9$\pm$0.5\\
T1149 & 10:27:35.88    & -43:52:51.1    &     22.2$\pm$0.1 & 22.1$\pm$0.1    &   0.0&     1.5$\pm$0.2  & 10.2$\pm$1.2\\
T1165 & 10:27:37.88    & -43:53:45.1    &     22.4$\pm$0.1  & 22.2$\pm$0.1   &   0.5 &    2.8$\pm$0.2  & 20.7$\pm$2.7 \\
\enddata
\tablenotetext{a}{\, Mean R$_{\rm eff}$ using the King30, King100, Moffat15 and Moffat25 profiles with ISHAPE routine}

\label{table1}
\end{deluxetable}


\begin{deluxetable}{lcccccccc}
\def\psn{\phs\phn}
\def\pnn{\phn\phn}
\tablecolumns{9}
\tablewidth{0pt}
\tablecaption{Indices}
\tablehead{
\colhead{ID} &\colhead{ $HHe$\tablenotemark{a} } &\colhead{ $K$\tablenotemark{a} } &\colhead{ $H8 $\tablenotemark{a} }  &\colhead{$H\gamma$\tablenotemark{b}}    &\colhead{ $Mgb5177$\tablenotemark{b}}    &\colhead{ $Fe5270$\tablenotemark{b}}    &\colhead{ $Fe5335$\tablenotemark{b}} &\colhead{log age} \\
\colhead{} & \colhead{ (\AA) } & \colhead{ (\AA) } & \colhead{ (\AA) }  & \colhead{( \AA)}& \colhead{( \AA)}& \colhead{( \AA)}& \colhead{( \AA)} & \colhead{( years)}
}
\startdata
T1127&   7.84$\pm $0.72 &   0.73$\pm $0.42    &6.96$\pm $0.74   &6.61$\pm $0.56      & 0.49$\pm$0.23  & 0.58$\pm$0.29   & 0.95$\pm$0.44    & 7.8$\pm $0.1  \\
T1149&   7.82$\pm $0.61&    0.53$\pm$0.37     &6.99$\pm$0.65    & 6.98$\pm$0.32      & 0.53$\pm$0.15  & 0.63$\pm$0.20   & 0.96$\pm$0.29     & 7.8$\pm $0.1\\  
T1165&   9.55$\pm $1.41 &   0.85$\pm$0.89    &8.59$\pm$1.56     & 7.93$\pm$0.76      & 0.77$\pm$0.36  & 0.87$\pm$0.45   & 1.20$\pm$0.64      & 8.3$\pm$0.2 \\ 
\enddata
\tablenotetext{a}{\,Indices definition by Schweizer \& Seitzer (1998).}
\tablenotetext{b}{\,Lick indices.}
\label{table:indices}
\end{deluxetable}


\begin{deluxetable}{lccccc}
\def\psn{\phs\phn}
\def\pnn{\phn\phn}
\tablecolumns{6}
\tablewidth{0pt}
\tablecaption{Cluster Positions and Radial Velocities}
\tablehead{
\colhead{} & \colhead{$r_{\rm p}$} &  \colhead{$r_{\rm p}$\tablenotemark{a}} &  
\colhead{\czhel} & \colhead{$\sigma_{cz}$} &
					\colhead{\dvlos\tablenotemark{b}} \\
\colhead{Cluster} & \colhead{(\arcsec)} & \colhead{(kpc)} &
\colhead{(\kms)} & \colhead{(\kms)} & \colhead{(\kms)}
}
\startdata
T1127   &  140 &  24.5 &  2750 & 19 &  \phn$-$67 \\
T1149   &  189 &  33.1 &  2663 & 16 &     $-$154 \\
T1165   &  150 &  26.3 &  2685 & 16 &     $-$132 \\
\enddata
\tablenotetext{a}{\,For $D = 36.1$ Mpc  ($H_0 = 70$ km s$^{-1}$ Mpc$^{-1}$).}
\tablenotetext{b}{\,Line-of-sight velocity relative to nucleus,
	$\dvlos = (\czhel-2818)/1.00940$\ \ (see text).}
\label{table4}
\end{deluxetable}

\end{document}